\documentstyle[preprint,aps]{revtex}

\begin{document}
\draft
%
%-----------------------------------TEXT------------------------------
%
\title
{Theory of Nonlinear Magneto-Optics}
\author{U. Pustogowa, T. A. Luce, {\bf W. H\"ubner}, and K. H. Bennemann}
\address{Institute for Theoretical Physics,
Freie Universit\"at Berlin, Arnimallee 14, D - 14195 Berlin, Germany}
\maketitle
%
%----------------------abstract---------------------------------------
%
\begin{abstract}
Nonlinear magneto-optics is a very sensitive fingerprint of the
electronic, magnetic, and atomic structure of surfaces, interfaces, and thin
ferromagnetic films.
Analyzing theoretically the nonlinear magneto-optical Kerr effect
for thin films of Fe(001) and at Fe surfaces
we demonstrate exemplarily how various electronic material properties of
ferromagnets , such as the $d$-band width, the magnetization, the substrate
lattice constant, and the film-thickness dependence can be extracted from
the calculated nonlinear Kerr spectra.
Furthermore, we show how the substrate $d$ electrons (Cu(001)) affect the
nonlinear Kerr spectra even without being themselves spin-polarized and
without film-substrate hybridization.
We show that the Kerr rotation angle in second harmonic generation is
enhanced by one to two orders of magnitude compared to the linear
Kerr angle
and how symmetry can be used to obtain the direction of magnetization in
thin films and at buried interfaces from nonlinear magneto-optics.
\end{abstract}
\pacs{75.30.Pd,78.20.Ls,73.20.At,75.50.Bb}

\newpage
%
%----------------------introduction------------------------------------
\section{INTRODUCTION}
Nonlinear magneto-optics for metals~\cite{Hue89,Pan} is a new expanding
field the significance of which results from extending electrodynamics to
the nonlinear case and from applications.
Particularly, an important task of theory was to extend the symmetry
analysis to an electronic theory~\cite{Hue94,Hue95}
since this allowed second harmonic generation (SHG) to be used as an
important new tool for the analysis of surface~\cite{Reif}, interface,
thin film, and multilayer~\cite{Rasprb,Rasprl} magnetism.
The surface sensitivity results since breakdown of inversion symmetry
is necessary for SHG.
Thus, it is possible to determine with the help of SHG
surface magnetic moments, magnetic anisotropy, domains, magnetostriction,
and, in particular, properties of buried interfaces.

Using an electronic theory (with a bandstructure input)
we calculate (i) the material specific nonlinear response function
$\chi^{(2)}(\omega ,{\bf M} ,\lambda_{s.o.},d)$, depending on magnetization
$M$, spin-orbit coupling $\lambda_{s.o.}$ and film thickness $d$,
(ii) the nonlinear Fresnel formula for the reflected nonlinear fields
$E(2\omega )$, and thus finally, (iii) the complex nonlinear Kerr
rotation angle $\Phi^{(2)}_{K}$.
The results, presented in the following demonstrate, in particular,
what can be learned (a) from an analysis of the frequency dependence
and (b) from the symmetry analysis and the Kerr rotation.
Some selected general results obtained previously are shown, but also
interesting new results indicating important substrate effects.
Remarkably, the calculation~\cite{raps} and recent
experiments~\cite{klaus,RasKo} yield that $\Phi_{K}^{(2)}$ is larger by
two orders of magnitude compared to the linear Kerr rotation.

\section{theory}

\subsection{Nonlinear susceptibility}

Using response theory
the nonlinear optical polarization is given by
	\begin{equation}
	P_{i}\;=\;\chi^{(2)}_{ijm}E_{j}(\omega )E_{l}(\omega ),
	\end{equation}
where $P_{i}$ is the i-th component of the polarizability and where the
susceptibility tensor $\chi^{(2)}_{ijl}$ is determined by~\cite{Hue89,Hue90}
\begin{eqnarray}
    &&\chi^{(2)}_{ijm}(2q_{\parallel },2\omega ,{\bf M})=
    \frac{e^{3}C}{\Omega} \frac{\lambda_{s.o.}}{\hbar \omega } \nonumber\\
    &&\times \sum_{\sigma }\sum_{{\bf k},l,l^{\prime },l^{\prime \prime }}
        \Bigg\{
    \langle {\bf k}+2{\bf q}_{\parallel },l^{\prime \prime }\sigma |i|
        {\bf k}l\sigma \rangle
      \langle {\bf k}l\sigma |j|{\bf k}+{\bf q}_{\parallel},l^{\prime }\sigma
        \rangle
      \langle {\bf k}+{\bf q}_{\parallel},l^{\prime }\sigma |m|
        {\bf k}+2{\bf q}_{\parallel },l^{\prime \prime }\sigma \rangle
     \nonumber\\
    &&\times  \frac
    {\frac{f(E_{{\bf k}+2{\bf q}_{\parallel },l^{\prime \prime }\sigma})-
           f(E_{{\bf k}+{\bf q}_{\parallel },l^{\prime }\sigma })}
         {E_{{\bf k}+2{\bf q}_{\parallel },l^{\prime \prime }\sigma }-
          E_{{\bf k}+{\bf q}_{\parallel },l^{\prime }\sigma }
          -\hbar \omega + i\hbar \alpha_{1}}
    -\frac{f(E_{{\bf k}+{\bf q}_{\parallel },l^{\prime }\sigma })-
           f(E_{{\bf k}l\sigma })}
         {E_{{\bf k}+{\bf q}_{\parallel },l^{\prime }\sigma }-
          E_{{\bf k}l\sigma }-\hbar \omega +i\hbar \alpha_{1}} }
    {E_{{\bf k}+2{\bf q}_{\parallel },l^{\prime \prime }\sigma }-
     E_{{\bf k}l\sigma }-2\hbar \omega +i2\hbar \alpha_{1}}
        \Bigg\} \; .
\label{gl1}
\end{eqnarray}
Here $E_{{\bf k}l\sigma }$ are the electronic energy levels resulting from
the bandstructure calculations, $\langle {\bf k}l\sigma |i|{\bf k}+
{\bf q}_{\parallel},l^{\prime }\sigma \rangle $ are the transition matrix
elements, which control the symmetry and thus the interface sensitivity.
If the matrix elements are treated as constants one has to introduce
a cut-off to guarantee the surface sensitivity.
This is done by the factor $C$~\cite{Pulmto}.
Typically, magnetic dipole effects are negligible if the electric dipole
contribution does not vanish.
$\lambda_{s.o.}$ is the key parameter for nonlinear magneto-optics.
In particular, for determing the Kerr effect it might be useful
to decompose $\chi^{(2)}$ in odd and even contributions
$\chi^{(2)}=\chi^{(2)}({\bf M})+\chi^{(2)}(M^{2})$.

For the application of the general theory thin films are of particular
interest.
Then, the layer-dependent contributions to $\chi^{(2)}$ reveal interesting
film properties and substrate effects.
For the thin film calculations we use $C=W_{{\bf k}+2{\bf q}_{\parallel},
l^{\prime \prime} \sigma }
W_{{\bf k}+{\bf q}_{\parallel },l^{\prime }\sigma } W_{{\bf k}l\sigma }$,
where $W_{\alpha }$ denotes the weight of the density of state
$|{\bf k}l\sigma \rangle $ in the Wigner-Seitz cell of the first
monolayer.
Note, due to the factor $C$ SHG will always result from the surface layer
or from the layer at the interface,
however the electronic structure of this layer depends on the film
thickness.
The bandstructure $E_{{\bf k}l\sigma }$ is calculated employing the
full-potential linear muffin-tin orbital method (FP-LMTO)~\cite{Meth} for
thin Fe(001) films within a symmetric slab calculation.

Results for one, three, five, and seven layers of Fe are presented in
Fig.~\ref{layer}.
The depth of the first minimum in the spectra scales linearly with the
magnetic moment of the top layer while the energy of this minimum
reflects the $d$-band width.
As can be seen from Fig.~\ref{lattice}, the nonlinear
spectra reveal the structural changes induced by different substrates or
temperatures.
In Fig.~\ref{semi} we compare the {\em ab initio} nonlinear Kerr spectrum
for the Fe(001) monolayer with tight-binding calculations.
{}From the results of Fig.~\ref{semi} it becomes obvious that it is necessary
to use {\em ab initio} calculations for the films if only bulk tight-binding
parameters are available.
However, tight-binding calculations may be adequate if one uses as input
parameters those determined from {\em ab initio} calculations for thin films.

We expect generally also interesting results for $\chi^{(2)}$ at
heterogeneous interfaces like Fe/Cu(001).
For example, it is of interest to find out how the substrate interferes
with the electronic structure of the thin film.
For this reason we performed calculations of the Fe/Cu bilayer.
Results are shown in Fig.~\ref{bilayer}.
The additional peak structure reveals the electronic influence even
of the nonmagnetic Cu on the nonlinear magneto-optic response of
thin films.
This might be of general significance also with respect to the analysis
of quantum well states (tuning of interface properties, $\lambda_{s.o.}$
in substrate, M in films).

{}From the results of Figs. 1 through 4 we learn that the nonlinear Kerr
effect is a sensitive probe of the magnetic and electronic structure
at interfaces.

\subsection{Symmetry and Kerr rotation}
That SHG is very symmetry sensitive is furthermore particular demonstrated
by its polarization dependence and by the remarkable enhancement of the
Kerr rotation\cite{raps}.
The latter important result was first theoretically derived by
Pustogowa {\em et al.}~\cite{raps} and later observed in
experiments by B\"ohmer {\em et al.}~\cite{klaus} and Koopmans
{\em et al.}~\cite{RasKo}.

First, we note that due to the symmetry sensitivity already SHG exhibits
a corresponding polarization dependence, which interestingly is very
different for transition and noble metals~\cite{Hue94}.
The latter results, since $s$ and $d$ electrons feel the breakdown of
inversion symmetry at the interface differently.
This symmetry dependence mainly results from the matrix elements in
$\chi^{(2)}$, see Eq.~(\ref{gl1}).
The different possible optical excitations contributing to $\chi^{(2)}$
for noble and transition metals are illustrated in Fig.~\ref{symm}.
Obviously, SHG yields interesting information on the electronic
structure even for $M$=0 and when no Kerr effect is present.
This area needs further analysis and might offer interesting new
results~\cite{Hue94}.

Extending the symmetry analysis to the magnetic case we determine the Kerr
rotations for different configurations and polarizations.
The general analysis can be performed by using for $s$ and $p$ polarization
(see~\cite{Hue95})
	\begin{equation}
	E^{(2\omega)}(\Phi,\varphi) \;=\;
	2i\left( \frac{\omega}{c}\right) \mid E^{(\omega)}_{0}\mid^{2}
	\times
	\left( \begin{array}{c}
	A_{p}F_{c}\cos\Phi\\
	A_{s}\sin\Phi\\
	A_{p}N^{2}F_{s}\cos\Phi
	\end{array} \right)
	\, \chi^{(2)}_{ilm} \,
	\left( \begin{array}{c}
	f_{c}^{2}t_{p}^{2}\cos^{2}\varphi\\
	t_{s}^{2}\sin^{2}\varphi\\
	f_{s}^{2}t_{p}^{2}\cos^{2}\varphi\\
	2f_{s}t_{p}t_{s}\cos\varphi\sin\varphi\\
	2f_{c}f_{s}t_{p}^{2}\cos^{2}\varphi\\
	2f_{c}t_{p}t_{s}\cos\varphi\sin\varphi
	\end{array} \right) \; ,
	\end{equation}
where in the longitudinal configuration the susceptibility tensor is given by
	\begin{equation}
	\chi^{(2)}_{ilm} \;=\;
	\left( \begin{array}{cccccc}
	0 & 0 & 0 & \mid\,0 & \chi^{(2)}_{xzx} & \chi^{(2)}_{xxy}\\
	\chi^{(2)}_{yxx} & \chi^{(2)}_{yyy} & \chi^{(2)}_{yzz} &
	\mid \chi^{(2)}_{yyz} & 0 & 0\\
	\chi^{(2)}_{zxx} & \chi^{(2)}_{zyy} & \chi^{(2)}_{zzz} &
	\mid\; \chi^{(2)}_{zyz} & 0 & 0
	\end{array}\right) \; ,
	\end{equation}
and in the polar configuration by
	\begin{equation}
	\chi^{(2)}_{ilm} \;=\;
	\left( \begin{array}{cccccc}
	0 & 0 & 0 & \mid\chi^{(2)}_{xyz} & \chi^{(2)}_{xzx} & 0\\
	0 & 0 & 0 & \mid\chi^{(2)}_{xzx} & -\chi^{(2)}_{xyz} & 0\\
	\chi^{(2)}_{zxx} & \chi^{(2)}_{zxx} & \chi^{(2)}_{zzz} &
	\mid\; 0 & 0 & 0
	\end{array}\right) \; .
	\end{equation}
Here, $\Phi $ and $\varphi $ denote the angles of polarization
of the reflected frequency doubled and of the incident light,
$A_{p,s}$ are the amplitudes,
$f_{c,s}$ and $F_{c,s}$ are the Fresnel coefficients, and
$t_{s,p}$ are the linear transmission coefficients for the
fundamental light.
Then, from
	\begin{equation}
	\phi _{K}^{(2)}\;\approx\;
	Re\frac{E^{(2\omega )}_{\varphi}(s-SH)}
	{E^{(2\omega )}_{\varphi}(p-SH)}
	\end{equation}
we determine the Kerr rotation for the polar and longitudinal configuration
and for $s$ and $p$ polarized incident light.
The following results are obtained:\\
For the longitudinal configuration ($M_{\parallel }$) and $p$ ($s$) polarized
incident light one gets
 	\begin{equation}
	\phi_{K,p}^{(2)M_{\parallel }} \;=\;
	Re\frac{E^{(2\omega )}_{p}(s-SH)} {E^{(2\omega )}_{p}(p-SH)} \;=\;
	Re \frac{a_{1}\chi^{(2)}_{yxx} + a_{2}\chi^{(2)}_{yzz}}
	{a_{3}\chi^{(2)}_{xzx} + a_{4}\chi^{(2)}_{zxx} +
	a_{5} \chi^{(2)}_{zzz}} \;,
	\label{eqlongp}
	\end{equation}
	\begin{equation}
	\phi_{K,s}^{(2)M_{\parallel }} \;=\;
	Re\frac{E^{(2\omega )}_{s}(s-SH)} {E^{(2\omega )}_{s}(p-SH)} \;=\;
	Re\frac{a_{6}\chi^{(2)}_{yyy}} {a_{7}\chi^{(2)}_{zyy}}.
	\label{eqlongs}
	\end{equation}
For the polar configuration ($M_{\perp }$) and $p$ ($s$) polarized
incident light one gets
	\begin{equation}
	\phi_{K,p}^{(2)M_{\perp }} \;=\;
	Re\frac{E^{(2\omega )}_{p}(s-SH)} {E^{(2\omega )}_{p}(p-SH)} \;=\;
	Re\frac{a_{8}\chi^{(2)}_{xyz}}
	{a_{9}\chi^{(2)}_{xxz} + a_{10}\chi^{(2)}_{zxx} +
	a_{11}\chi^{(2)}_{zzz}} \; ,
	\label{eqpolp}
	\end{equation}
	\begin{equation}
	\phi_{K,s}^{(2)M_{\perp }} \;=\;
	Re\frac{E^{(2\omega )}_{s}(s-SH)} {E^{(2\omega )}_{s}(p-SH)} \;=\;
	Re\frac {0}{a_{12}\chi^{(2)}_{zxx}}\;.
	\label{eqpols}
	\end{equation}
The coefficients $a_{1} \ldots a_{12}$ contain explicitly the whole
information about the Kerr geometry (directions of light incidence and
polarization, crystal magnetization) and about the linear transmission
and reflection (see~\cite{Hue95}).

Summarizing the results, we find that generally
$\phi_{K}^{(2)M_{\perp }} > \phi_{K}^{(2)M_{\parallel }}$ and
$\phi_{K,s}^{(2)} > \phi_{K,p}^{(2)}$, which is in agreement with the
experiment by Koopmans {\em et al.}~\cite{RasKo}.
Numerical results of our theory using as input parameters the phase ratio
of the various different tensor elements
from Ni and linear complex indices of refraction are presented in
Fig.~\ref{angle}.
These results demonstrate clearly that the nonlinear Kerr angle depends
sensitively on the direction of the magnetization and the incident beam,
as well as on the light polarization.
Depending on the experimental conditions the nonlinear Kerr angle might
become even as large as 90$^{\circ }$.
Moreover, also the ellipticity will yield correspondingly large magnetic
effects.
{}From these results we conclude that the enhanced nonlinear Kerr
rotation allows for the determination of easy axis and canted-spin
configurations due to large magnetic contrast of different interfaces.

This selective discussion demonstrates the usefulness of nonlinear
magneto-optics as a promising spectroscopy.
Our results show that SHG is a new sensitive tool for determining
magnetic interface properties such as the interplay of structure and
magnetism, magnetic anisotropy, and magnetic reorientation.
In contrast to linear optics, which probes characteristic film-averaged
features, the nonlinear Kerr effect originates essentially from the
surface and interfaces and thus allows the analysis of buried interfaces.

For future studies on the nonlinear magneto-optical Kerr effect
it will be interesting to analyze magnetic anisotropy and easy axis,
magnetic structure and domains, multilayers, and magnetostriction effects.
Nonlinear optics at interfaces is a bulk-background-free technique
and therefore, should yield a pronounced visibility of quantum well states
and enhanced nonlinear paramagnetic Kerr oscillations in the presence
of an external field.

An important future development of the nonlinear Kerr effect might
result from femto-second dynamics which is able to follow the charge and
spin dynamics of metallic interfaces in real time without involving
the lattice.

%
%-------------------------references-----------------------------------
\newpage
\newpage
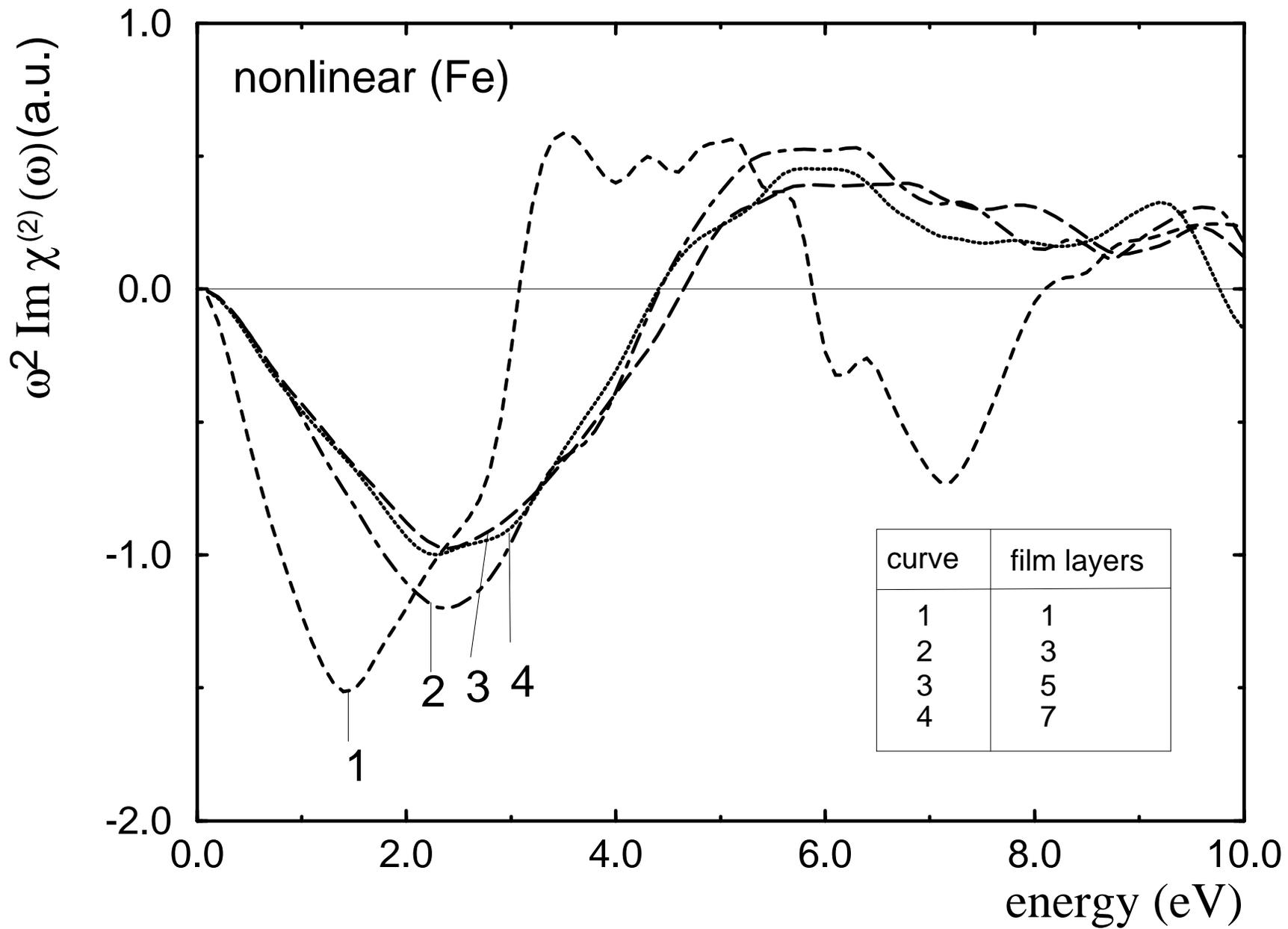
\begin{figure}
\caption[]{{\em Ab initio} calculated nonlinear magneto-optical Kerr spectra
of Fe using Eq. (2) for a monolayer (dashed curve - 1), and
films with 3 layers (dashed-dotted curve - 2), 5 layers (long-dashed - 3),
and 7 atomic layers (dotted curve - 4).
The second harmonic response results from the first atomic layer.
This SH response is obtained by projecting the wave functions
to the first atomic layer yielding the factor $C$.}
\label{layer}
\end{figure}
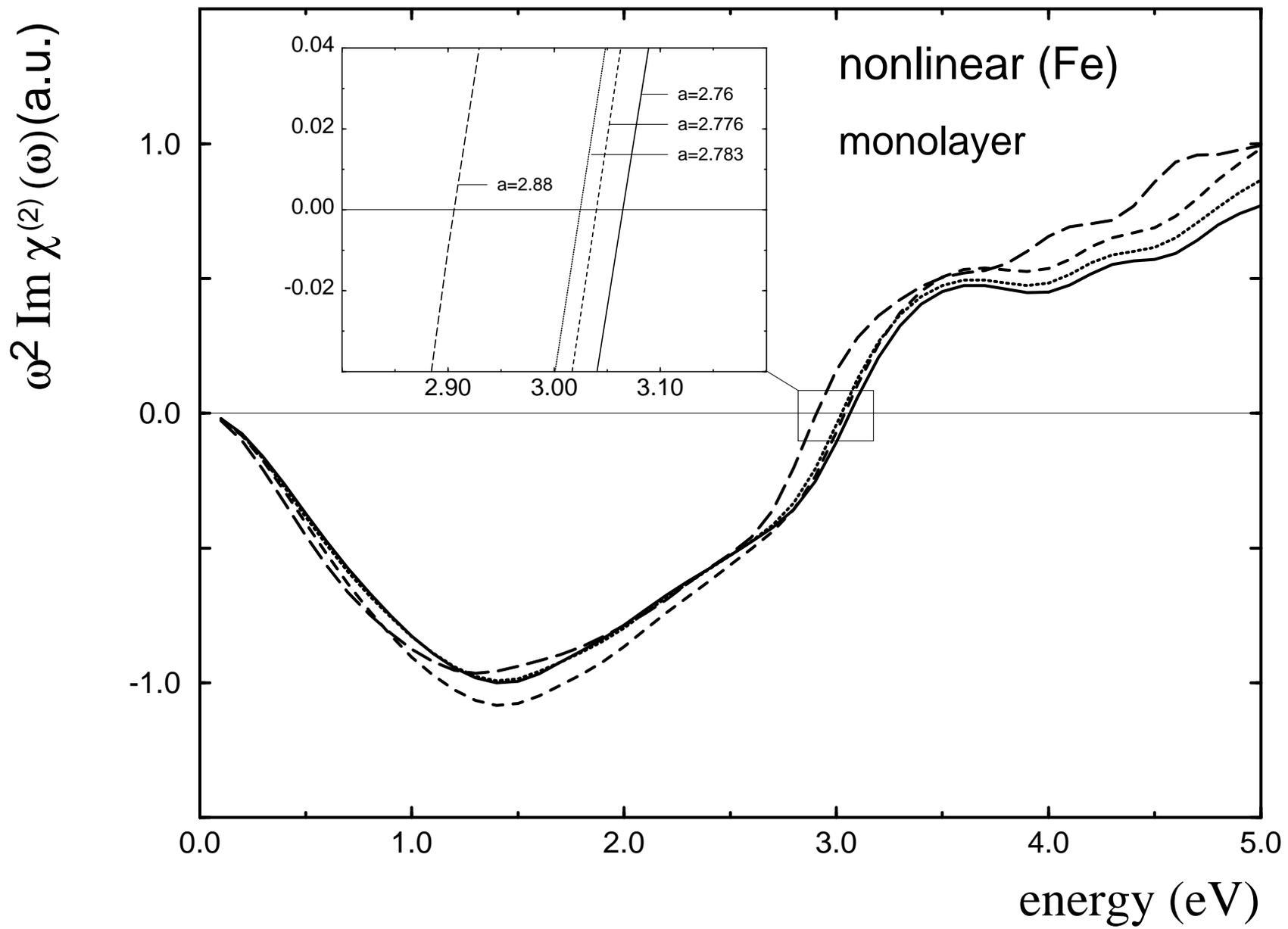
\begin{figure}
\caption[]{Film-lattice-constant dependence of {\em ab initio} calculated
nonlinear Kerr spectra of a Fe monolayer.
The solid curve refers to the bulk {\em bcc} Fe lattice constant
$a$=2.76 \AA, the dashed curve to $a$=2.776 \AA $\,$(bulk Au), and the dotted
curve to $a$=2.783 \AA $\,$(bulk Ag).
The long-dashed curve refers to the experimental $a$=2.88 \AA $\,$for Fe.
Choosing lattice constants $a$ refering to Au and Ag should simulate
substrate effects.
The inset shows for an enhanced scale the effects of different lattice
constants for the zero of Im $\chi^{(2)} $ at $\hbar \omega \approx $ 3 eV
and indicate structural effects which should be observable.}
\label{lattice}
\end{figure}
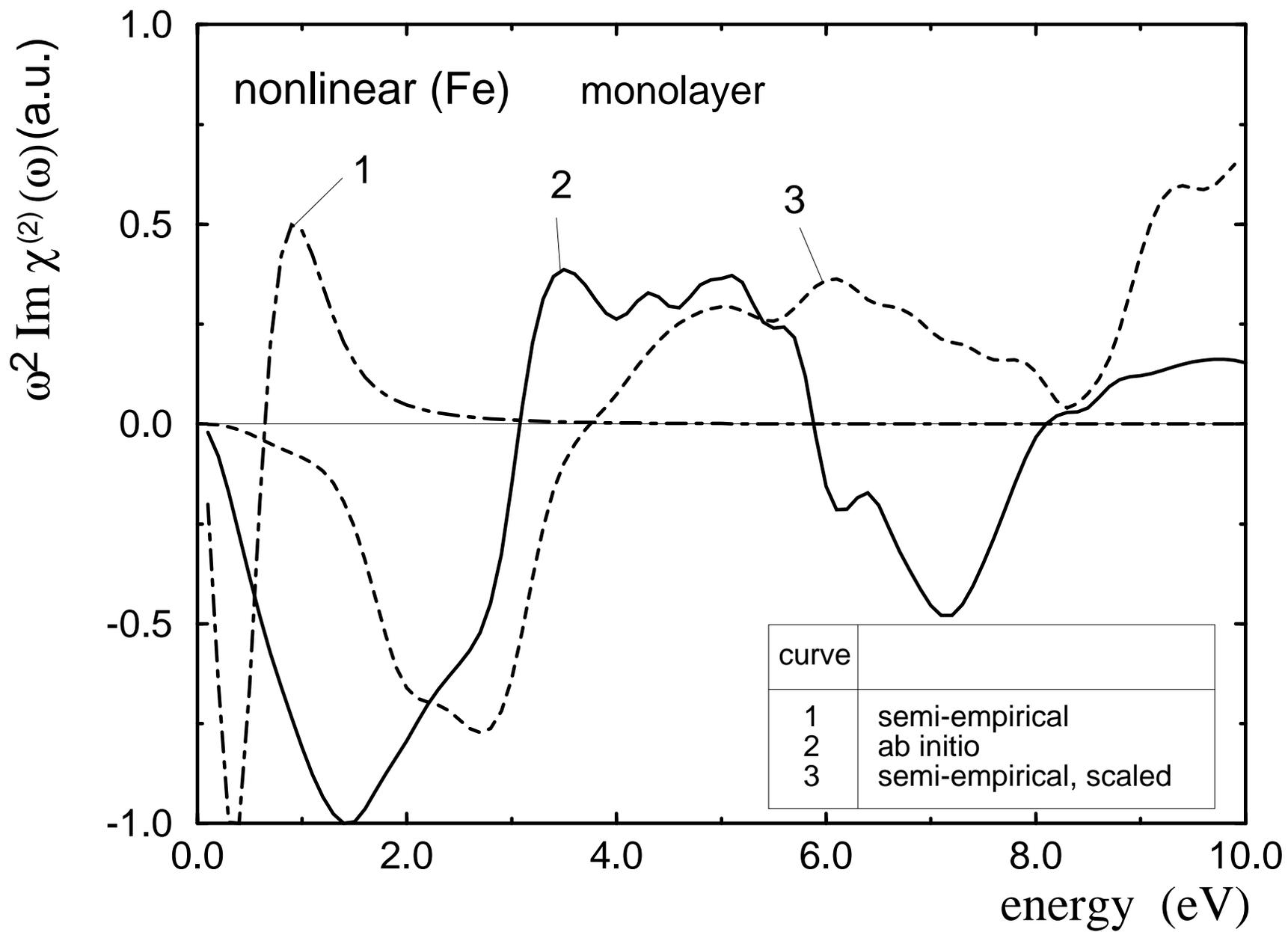
\begin{figure}
\caption[]{Comparison of {\em ab initio} (solid line - 2) and semi-empirical
calculations of the nonlinear magneto-optical susceptibilities for a Fe
monolayer.
Semi-empirical calculations are performed for a two-dimensional atomic
configuration (dashed-dotted curve - 1) and for a three-dimensional one with
reduced hopping parameters (dashed curve - 3) containing the reduced
coordination number of a monolayer compared to the bulk.}
\label{semi}
\end{figure}
\begin{figure}
\caption[]{Nonlinear magneto-optical Kerr-spectrum of a Fe/Cu bilayer system
without (solid curve) and with (dashed curve) hybridization between the
Fe and Cu layers.
Note, the spectrum look rather different from the corresponding one for
freestanding monolayers indicating a strong effect of Cu on the optical
transitions.}
\label{bilayer}
\end{figure}
\begin{figure}
\caption{Illustration of second harmonic generation from noble metals
[(a) and (b)] and transition metals [(c) and (d)]. For noble metals,
in case (a) no $d$ electrons can be optically excited, which in contrast is
possible in case (b). In case (c) for transition metals, predominantly $d$
electrons contribute to the SHG yield, whereas in case (d) the excitation
starts from the $s$ band.
Note, the cases (a) and (c) refer to low-frequency excitations while (b)
and (d) correspond to high-frequency excitation.}
\label{symm}
\end{figure}
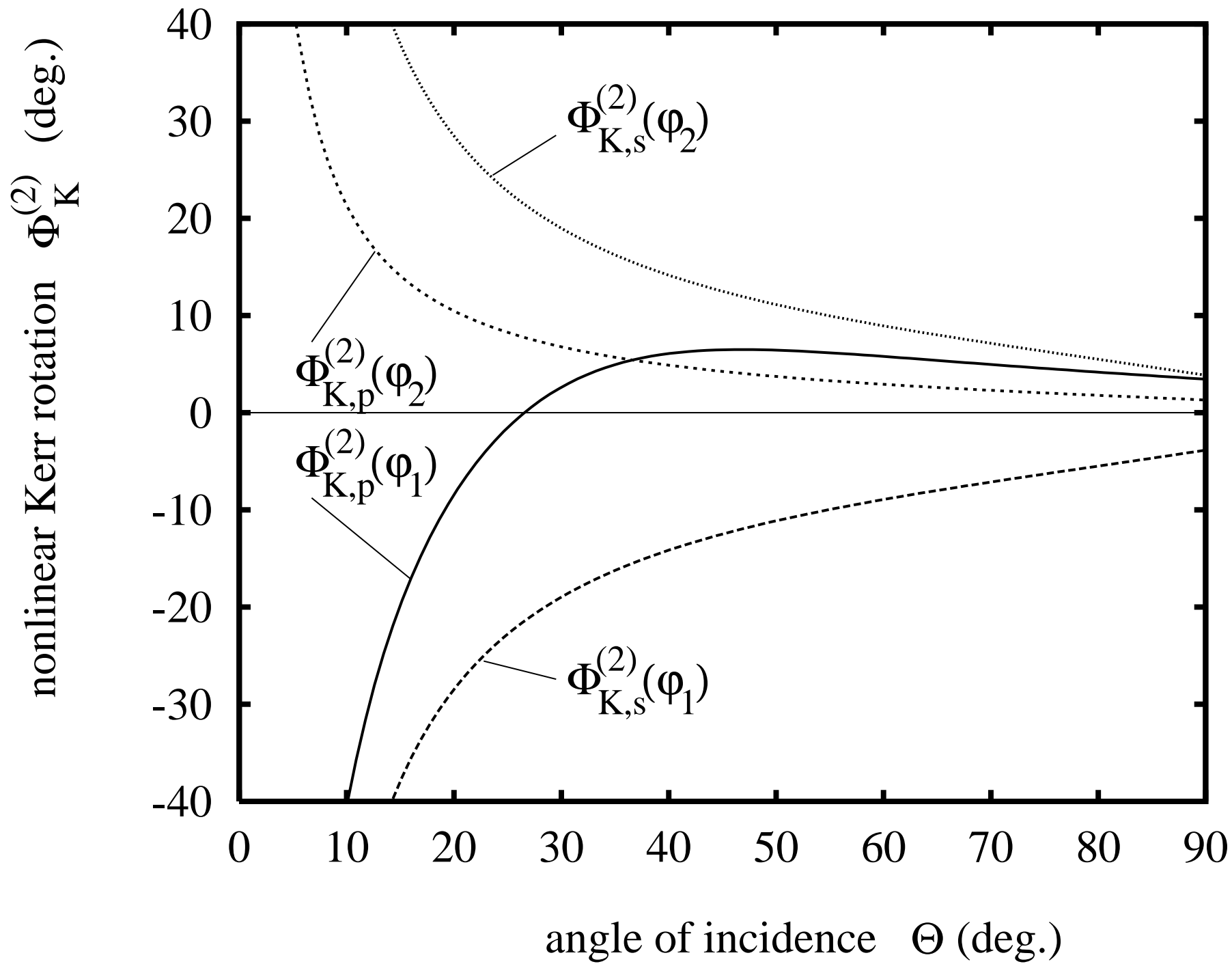
\begin{figure}
\caption{Nonlinear Kerr rotation angles for $p$ polarized incident light
$\phi^{(2)}_{K,p}$ (full and short-dashed curves) and for $s$
polarized incident light $\phi^{(2)}_{K,s}$ (long-dashed and dotted
curves) for Fe at 770 nm as a function of the angle of incidence $\theta $
in the longitudinal Kerr configuration.
The relative phase between $\chi^{(2)}_{zxx}\;=\;\chi^{(2)}_{zyy}$ and
$\chi^{(2)}_{zzz}$ is $\varphi_{1}$=0.505$\pi$ in the full and long-dashed
curves and $\varphi_{2}$=1.505$\pi$ in the short-dashed and dotted curves.}
\label{angle}
\end{figure}
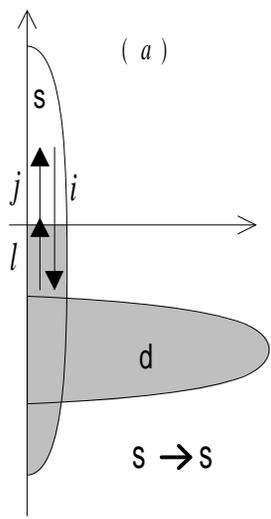
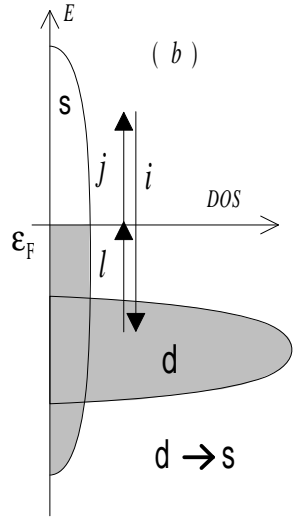
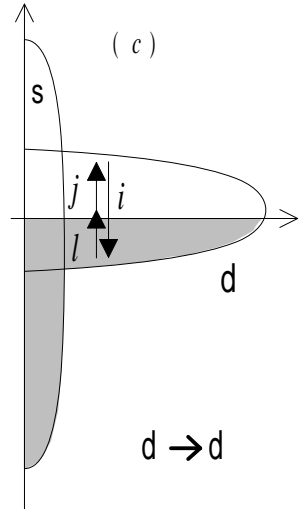
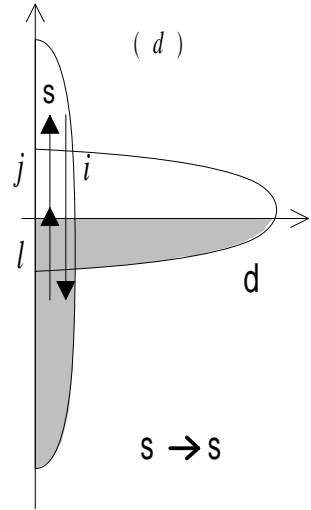
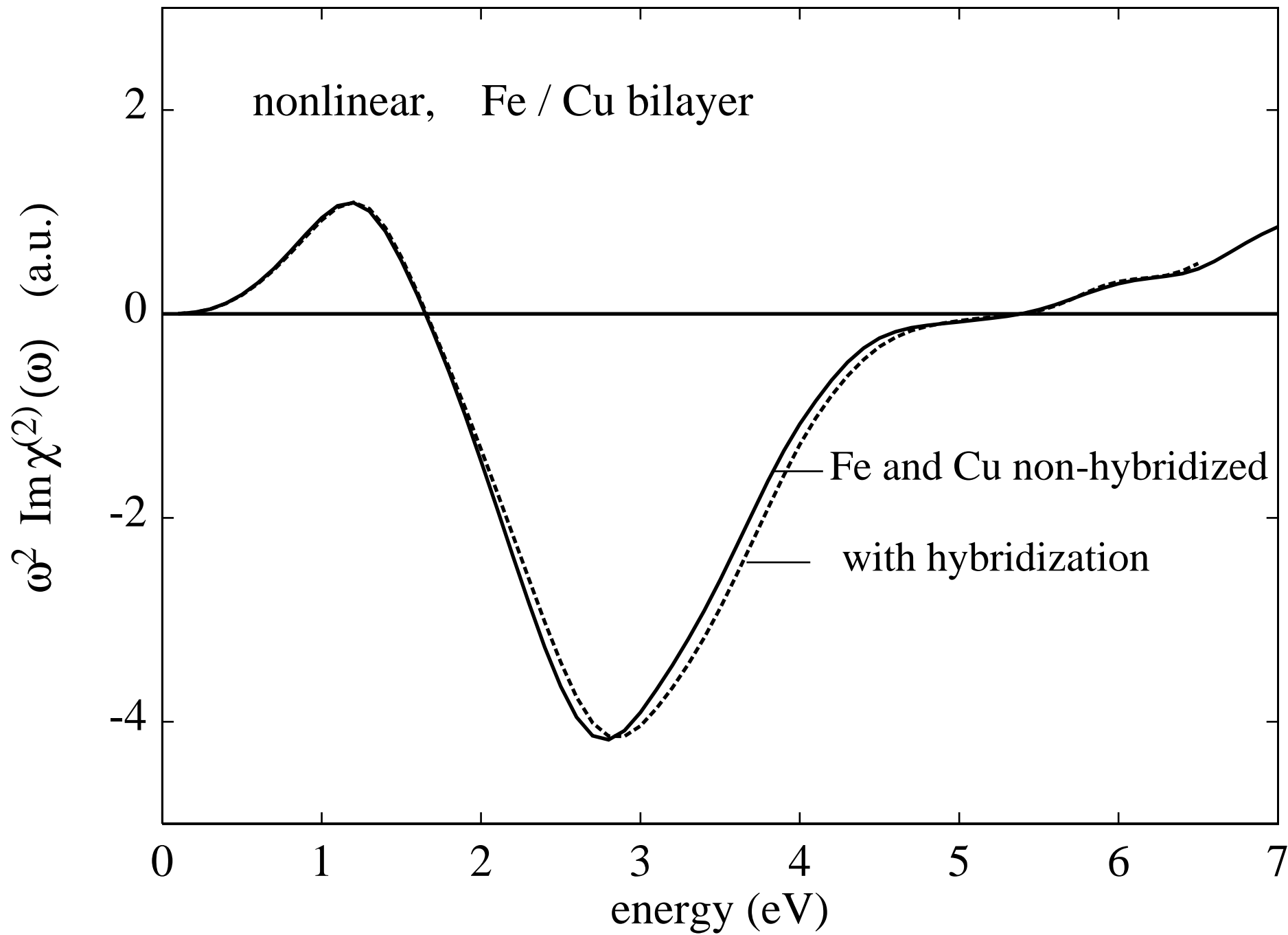
\end{document}